\documentstyle[aps,prc,twocolumn]{revtex}
\begin{document}
\title{ Model Analysis of Time Reversal Symmetry Test\\
         in the CALTECH $^{57}$Fe $\gamma$-transition experiment }
\draft

\author{
                                 Michael Beyer}
\address{            Institut f\"ur Theoretische Kernphysik,\\
               Universit\"at Bonn, Nussallee 14-16, 5300 Bonn, FRG}
\date{\today}

\maketitle
\begin{abstract}
The CALTECH $\gamma$-transition experiment testing time reversal
symmetry via the E2/M1 mulipole mixing ratio of the 122 keV
$\gamma$-line in $^{57}$Fe has already been performed in 1977. Extending
an earlier analysis in terms of an effective one-body potential, this
experiment is now analyzed in terms of effective one boson exchange
T-odd P-even nucleon nucleon potentials. Within the model space
considered for the $^{57}$Fe nucleus no contribution from isovector
$\rho$-type exchange is possible. The bound on the coupling strength
$\phi_A$ from effective short range axial-vector type exchange induced
by the experimental bound on $\sin\eta$ leads to $\phi_A \leq 10^{-2}$.
\end{abstract} 

\narrowtext

\section{introduction}
The existence of CP-violation is well established through decays in the
$K_L$, $K_S$ system \cite{chr64,PDG}, and could be accomodated by the
Kobayashi-Maskava model\cite{KM}. Other models of CP-violation implied
via the weak sector are e.g. the Weinberg-Higgs model \cite{wei76} or
left-right symmetric models including nonstandard right handed currents
\cite{rlsym}. The milliweak model predicts $\epsilon'/\epsilon=0$
\cite{wol64}. Experimental results on $\epsilon'/\epsilon$ are
debated\cite{wol92}. CP/T-symmetry could also be broken by millistrong,
strong or electromagnetic forces. Although the breaking by millistrong
and electromagnetic forces have a somewhat weaker stand, they cannot be
ruled out completely; for details see ref. \cite{wol86}. In addition,
the $\theta$-term of QCD could also provide a source of
CP-violation\cite{theta}.

However, besides in the $K$-system no further evidence of CP-violation
and/or T-violation has been found to date.

One of the classical tests on time reversal symmetry is the search of
T-odd correlations in $\gamma$-transitions of nuclei \cite{hen59,boe68}.
Many experiments have been conducted \cite{gamma,che77,sac87}, but lead
only to upper bounds. The highest accuracy in this class of experiments
has probably been reached by the $\gamma$-transition in $^{57}$Fe done
at CALTECH in 1977 \cite{che77}, which will be considered in the
following.

Additional tests on time reversal symmetry are the search for electric
dipole moments of the neutron \cite{edmn}, the atom \cite{edma} and the
electron \cite{edme}, T-odd correlations in $\beta$-decay \cite{beta},
tests of detailed balance, e.g. via compound-nucleus reactions
\cite{db}, polarization experiments \cite{pol}, and transmission of
thermal neutrons through nuclei \cite{ntrans}. Earlier references may be
found in \cite{sac87}.

New experiments involving nuclei have been suggested. Among them are
high precision tests using $\gamma$-decays\cite{w182}, neutron
transmission \cite{LANCE}, and proton deuteron forward scattering
\cite{COSY,bey93}.

Since the basic mechanism of possible CP/T-violation is not known, in
principle each type of experiment is unique. In order to compare the
bounds reached by different experiments, in particular those expected
from future tests, some model assumptions are necessary, at least to
treat the dynamic behavior of the system in question. Due to the
moderate energies involved in most of the nuclear experiments, effective
hadronic degrees of freedom, viz. mesons and nucleons, may be a
reasonable choice to model T-violation in a nuclear environment. An
approach along this line has been quite useful in the treatment of
parity violation in nuclei \cite{DDH}. Though not fundamental, hadronic
degrees of freedom have not only been very successful, but belong to the
best ``realistic'' describtions of the strong NN interaction in the
energy regime also relevant here, see e.g. \cite{bonn}. Nevertheless, an
appropriate description in terms of the models of CP/T-violation given
above is certainly desireable, see e.g. \cite{khr91}, but would go
beyond the scope of the present approach.

\section{effective CP/T-violating hadronic interactions}

A simplification for complex nuclei is possible using effective one-body
potentials. This concept has
been quite useful in the context of P-violation
\cite{mic64} and has also been used to model T-violating effects
\cite{bli73,bey89,gri91}. In case of parity conserving T-violation that
is relevant for the experiment in question a general effective one-body
potential may be written as (density dependence neglected)
\cite{bli73,bey89,her66}
\begin{equation}
U_T = \frac{G_T}{2} \sum_i
      \left( {\bf p}_i \cdot {\bf \hat{r}}_i
      + {\bf \hat{r}}_i\cdot{\bf p}_i\right)
\label{UT}
\end{equation}
with unit vector ${\bf \hat{r}}_i $ and coupling constant $G_T$. For the
$^{57}$Fe nucleus the bound induced by the CALTECH experiment
\cite{che77} has been found to be $|G_T|\leq 5\times 10^{-6}$ \cite{bey89}.

However $G_T$ contains still some information related to the specific
nucleus. To make this obvious, the effective one body potential $U_T$
given above may be related to a more basic nucleon-nucleon (NN)
potential $V_T$ via a Hatree-Fock potential, viz.
\begin{eqnarray}
\lefteqn{{\left\langle\,{\psi_\alpha}\,\left|\,{U_T}\,
            \right|\,{\psi_\beta}\,\right\rangle}=}\nonumber\\
&&\quad\sum_{\chi<F}\left(
{\left\langle\,{\psi_\alpha\chi}\,\left|\,{V_T}\,
            \right|\,{\psi_\beta\chi}\,\right\rangle}
-{\left\langle\,{\psi_\alpha\chi}\,\left|\,{V_T}\,
            \right|\,{\chi\psi_\beta}\,\right\rangle}
\right)
\label{HF}
\end{eqnarray}
where $\chi$ are occupied levels only, and the first sum on the
right hand side is the direct, the second the exchange term.
This approach has been used by Grimlet and
Vogel, investigating  T- and P-violating pion exchange
potentials \cite{gri91}.

The general structure of T-violating NN potentials has first been given
by Herczeg \cite{her66}. A field theoretical approach that lead to one
boson exchange T-violating P-conserving potentials have been considered
by Sudarshan \cite{sud68}, Bryan and Gersten \cite{bry71}, and by
Simonius and Wyler, who also analyzed polarization observables
\cite{sim75,sim77}. Huffman has constructed effective NN potentials from
T-violating electromagnetic currents \cite{huf70}.

Since the $\gamma$-transition experiment considered here is
P-conserving, in the following I consider only this type of T-violation.
Note however, that the question whether the standard  model  alone
provides  any P-conserving T-violating interaction on tree level,  is
presently  under  discussion \cite{her88,privat}.

Parametrizing CP/T-violation in terms of one boson exchange leads to
vector and axial vector exchanges, only. A long range $\pi$-meson
exchange requires simultanious violation of P and T \cite{sim75}. The
($\rho$-meson) vector exchange, with a C-violating (and hence
T-violating) isospin dependence reads up to order {\bf p}$^2/m^2_p$
\cite{sim75,sim77}:
\begin{eqnarray}
V^\rho_T&=&\phi_\rho\kappa_\rho\frac{g^2_{\rho NN}}{4\pi}
          \frac{m_\rho^3}{8m^2_p} F(m_\rho r)     \nonumber\\
&&          (\bbox{\sigma}_1\!-\!\bbox{\sigma}_2)
           \cdot{\bf r}\!\times\!{\bf p}
          (\bbox{\tau}_1\!\times\!\bbox{\tau}_2)_0 + h.c.
\label{rho}
\end{eqnarray}
Here and in the following ${\bf r}=({\bf r}_1-{\bf r}_2)$,
${\bf p} = \frac{1}{2}({\bf p}_1-{\bf p}_2)$, and
\begin{equation}
F(x) = \frac{e^{-x}}{x^3}(1+x).
\end{equation}
The strength $\phi_\rho=g^T_{\rho NN}/g_{\rho NN}$ denotes the strength
of the T-violating  potential relative to the T-conserving one. The
other parameters  are $\rho$-meson mass $m_\rho$, proton mass $m_p$, and
the ratio of tensor to vector strength $\kappa_\rho=f_{\rho NN}/g_{\rho
NN}$.

The axial vector exchange  \cite{sud68,sim75,sim77} reflects the
different behavior of axial vector and pseudo  tensor interactions
under time reversal symmetry. Up to order {\bf p}$^2/m^2_p$  it is given by
\begin{eqnarray}
V^A_T&=&\phi_A \frac{g^2_{ANN}}{4\pi}
            \frac{m_A^3}{8m^2_p}F(m_Ar)\nonumber\\
&&          ( \bbox{\sigma}_1\!\cdot\!{\bf p}
            \bbox{\sigma}_2\!\cdot\!{\bf r}
           +\bbox{\sigma}_2\!\cdot\!{\bf p}
            \bbox{\sigma}_1\!\cdot\!{\bf r}
           -\bbox{\sigma}_1\!\cdot\!\bbox{\sigma}_2
            {\bf p}\!\cdot\!{\bf r}) + h.c.
\label{axial}
\end{eqnarray}
with $m_A=1.26$GeV the axial vector  meson  mass.  The  isospin
dependence is  not restricted and may be isoscalar, isovector or
isotensor. The coupling strength $g_{ANN}$ is  not  well  know
empirically  from  NN potentials.  One  may  choose $g^2_{ANN}/4\pi=3.8$
\cite{sud68,bey93} for an isoscalar $\bbox{\tau}_1 \cdot \bbox{
\tau}_2$-dependence.

\section{the $^{57}$F\lowercase{e} experiment}

The experiment considered is sensitive to $\sin\eta$, the T-odd part in
the multipole mixing ratio $\delta$ of the mixed $\gamma$-transition of
nuclei \cite{hen59}. The ratio $\delta$ of the reduced matrix elements
is defined by \cite{bie53}
\begin{equation}
\delta = \frac
{\left\langle\,{J_f}\,\left\|\,{T^{\pi'}_{L'}}\,
            \right\|\,{J_i}\,\right\rangle}
{\left\langle\,{J_f}\,\left\|\,{T^\pi_L}\,
            \right\|\,{J_i}\,\right\rangle}
=|\delta|(\cos\eta+i\sin\eta)
\end{equation}
with $L'=L+1, \pi'\neq\pi$. The measured ratio has been the E2/M1
(=$\pi' L'/\pi L$) mixing in the 122.1 keV line from the $\frac{5}{2}^-$ to
the $\frac{3}{2}^-$ state of $^{57}$Fe. The correlation used to measure an
upper bound on $\sin\eta$ is $({\bf J}\!\cdot\! {\bf q}\!\times\! {\bf
E}) ({\bf J}\!\cdot\!{\bf q})({\bf J}\!\cdot\!{\bf E})$ with {\bf J}
quantization axis of the initial state, {\bf E} electric field vector,
and {\bf q} photon direction. For more details see ref. \cite{che77}.
The value quoted is \cite{che77}
\begin{equation}
|\sin\eta| = (3.1\pm 6.5)\times 10^{-4}.
\label{sinexp}
\end{equation}
The sources adding up to a finite $\sin\eta$ may not only be T-violating
interactions, but also due to final state interactions that results in a
phase shift $\xi$. These have been considered for the experiment in
question and found smaller than the experimental accuracy \cite{che77}.
Writing $\eta$ as ($\eta_0 = 0$ or $\pi$)
\begin{equation}
\eta = \eta_0
+\varepsilon_{E2}-\varepsilon_{M1}+\xi
\hspace{1cm}
\varepsilon_{E2},\varepsilon_{M1}, \xi \ll 1
\end{equation}
the CP/T-violating contributions $\varepsilon$ to the E2/M1 mixing ratio
in first order pertubation theory are given by \cite{bli73,bey89}
\begin{eqnarray}
\lefteqn{i(\varepsilon_{E2} -\varepsilon_{M1}) =
\frac{\left\langle{J_f}\left\|{E2_T}
            \right\|{J_i}\right\rangle}
{\left\langle{J_f}\left\|{E2}
            \right\|{J_i}\right\rangle}
-\frac{\left\langle{J_f}\left\|{M1_T}\,
            \right\|{J_i}\right\rangle}
{\left\langle{J_f}\left\|{M1}
            \right\|{J_i}\right\rangle}}\nonumber\\
&+&\sum_n
\frac {
{\left\langle{J_f}\left|{V_T}
            \right|{nJ_f}\right\rangle}}{E_f-E_n}
\left(\frac{\left\langle{nJ_f}\left\|{E2}
            \right\|{J_i}\right\rangle}
{\left\langle{J_f}\left\|{E2}
            \right\|{J_i}\right\rangle}
-\frac{\left\langle{nJ_f}\left\|{M1}
            \right\|{J_i}\right\rangle}
{\left\langle{J_f}\left\|{M1}
            \right\|{J_i}\right\rangle}\right)\nonumber\\
&+&\sum_n
\frac {
{\left\langle{nJ_i}\left|{V_T}
            \right|{J_i}\right\rangle}}{E_i-E_n}
\left(\frac{\left\langle{J_f}\left\|{E2}
            \right\|{nJ_i}\right\rangle}
{\left\langle{J_f}\left\|{E2}
            \right\|{J_i}\right\rangle}
-\frac{\left\langle{J_f}\left\|{M1}
            \right\|{nJ_i}\right\rangle}
{\left\langle{J_f}\left\|{M1}
            \right\|{J_i}\right\rangle}\right)
\label{eps}
\end{eqnarray}
The first two contributions, $E2_T$, $M1_T$ refer to a generic one-body
T-odd part in the electromagnetic multipole operator, and will be
neglected in the following. It could arise from a direct CP/T-violating
part in the electromagnetic interaction.

The two body contribution results from an admixture of a CP/T-violating
potential $V_T$ as e.g. given above. It will be evaluated in a shell
model for the $^{57}$Fe nucleus in the follwing.

The model space for $^{57}$Fe is restricted to three valence neutron
particle states in the ($2p_{1/2}, 2p_{3/2}, 1f_{5/2}$) and two
valence proton hole states in the $1f_{7/2}$ shell. The other nucleons
are in a closed ($T=0$, $J=0$) $^{56}$Ni shell. Extensions of the model
space have not been found sufficient for low-lying states. Relevant 4p3h
states are expected to be not lower than 3MeV \cite{kos87}. The surface
delta interaction (SDI) has been used as residual interaction. This has
been suggested in ref. \cite{shell,ven80} and proven to give a
reasonable well overall describtion of the spectroscopic data of the $A=
56, 57, 58$ Fe isotops \cite{ven80}. The energy spectrum, electromagnetic
transition rates, static properties as the magnetic and quadrupole
moments are sufficiently well described \cite{ven80}, see also
\cite{bey89}.

The model given above has been used earlier in the context of
T-violation with an estimated uncertainty on $G_T$ of 50\% or so
\cite{bey89}. This uncertainty reflects the comparison of different
(finite) sets of basis states, viz. harmonic oscillator vs. Saxon-Wood
single particle wave functions. A second source are different residual
interactions, viz. SDI vs. Johnstone-Benson, the later using a set of
fitted matrix elements \cite{JB}. Since the present calculation uses
``elementary'' two body interactions the bounds on the T-odd coupling
constant $\phi$ induced by eq.(\ref{sinexp}) may have additional
uncertainties due to the (necessary) restriction of the model space. The
same reason leads to the concept of effective charges of $e_p=2$ for the
proton and $e_n=1$ for the neutron \cite{shell,ven80} and has also been
used here to evaluate the $\gamma$-transition reduced matrix elements.

The nuclear wave functions emerging from the shell model approach lack
the short range repulsive NN-correlations. These are particular
important, since the T-violating P-conserving NN-potentials are
essentially short range. The short range correlation has been
accomodated by multiplying the two nucleon wave functions, necessary to
evaluate the T-odd admixture in eq.(\ref{eps}) by \cite{short}
\begin{equation}
f(r) = 1 - e^{-ar^2}(1-br^2)
\label{SR}
\end{equation}
with $a=1.1$fm$^{-2}$ and $b=0.68$fm$^{-2}$. This results in a
replacement of two body matrix elements according to
\begin{equation}
{\left\langle\,{\psi_\alpha'\psi_\beta'}\,\left|\,{V_T}\,
            \right|\,{\psi_\alpha\psi_\beta}\,\right\rangle}
\rightarrow
{\left\langle\,{\psi_\alpha'\psi_\beta'}\,\left|\,{fV_Tf}\,
            \right|\,{\psi_\alpha\psi_\beta}\,\right\rangle}.
\end{equation}

\section{results and discussion}

Evaluation of ${\left\langle\,{nJ}\,\left |\,{V_T}\,\right|
\,{J}\,\right\rangle}$ in eq.(\ref{eps}) leads to contributions denoted
as particle-particle, hole-particle, hole-hole and particle(hole)-core
parts \cite{BM}. The later has already been given in eq.(\ref{HF}).
However, neither of the above potentials, eqs.(\ref{rho},\ref{axial}),
give rise to a particle(hole)-core type of effective potential. The
isospin dependence of the $\rho$-exchange, eq.(\ref{rho}), causes the
direct part to vanish, the spin depence leads to vanishing exchange
contribution of eq.(\ref{HF}). The direct part of the axial exchange,
eq.(\ref{axial}), particle(hole)-core contribution vanishes due to the
spin dependence and the exchange part cancels, since the density of
occupied (single particle shell model) states is local, viz.
\begin{equation}
\sum_\chi
{\left\langle{\bf x'}\!\right.\left|{\chi}\right\rangle}
{\left\langle{\chi}\!\right.\left|{\bf x}\right\rangle}
=\delta({\bf x}'-{\bf x})\rho({\bf x}).
\end{equation}

Unfortunately, all the valence contribution of the $\rho$-type potential
also vanish in the model space chosen for $^{57}$Fe. This is due to the
isovector character of the T-odd potential for both, the direct and
exchange part of eq.(\ref{HF}). Note, that neutron particles and proton
holes have the same isospin projection. Extention of the model space,
e.g. excitation of proton particle-hole pair, 4p3h states, might lead to
finite results, however, but may be suppressed due to excitation energy.

The axial exchange leads to valence contributions to $\sin\eta$ for scalar
and tensor parts. They are approximately of the same size. The
contribution from admixture to the $\frac{3}{2}^-$ state, the first sum
in eq.(\ref{eps}), is one order of magnitude larger than the
contribution due to the admixture to the $\frac{5}{2}^-$ state. The result
is
\begin{equation}
\sin\eta = 3.14\times 10^{-2}\phi_A
\end{equation}
which leads to a bound on $\phi_A$ implied by eq.(\ref{sinexp}) of
\begin{equation}
\phi_A = (1\pm 2)\times 10^{-2}
\label{bound}
\end{equation}
with some theoretical uncertainty as mentioned above. Note, that
neglecting the short range correlation, eq.(\ref{SR}), would result in a
much stronger (but fake) limit on the coupling strength by roughly two
orders of magnitude.

The bound on the model strength $\phi_A$ is less ambigiuos concerning
the nuclear structure than the bound in terms of effective one-body
potentials. It may be compared to other experiments analyzed with the
same model NN interaction, and eventually related to more fundamental
T-violating Langrangians in terms of quark and gluonic degrees of
freedom \cite{khr91}.

The bound given in eq.(\ref{bound}) may be compared to similar bounds,
estimated from electric dipole moments. Since a finite electric dipole
moment requires simultanious P- and T-violation one may assume
P-violation to be of the order of the weak interaction, viz.
$G_Fm_\pi^2=2\times 10^{-7}$ ($m_\pi$ mass of the pion). Implications on
a generic T-odd P-even meson nucleon coupling strength may then be found
from the upper limit on the neutron electric dipole moment. An
``educated guess'' gives $\phi_A< 10^{-4}$ \cite{her88,khr91}. However,
no dynamical consideration has been done in deriving this bound. For
details see ref. \cite{khr91,her88}.

Compared to the bound on the effective one-body potential eq.(\ref{UT}),
the bound given in eq.(\ref{bound}) is almost 4 orders of magnitudes
weaker. However, since only the valence nucleons contribute to the
T-violating effect, the ``nuclear structure factor'' is of the same
magnitude as for other few particle systems, compare \cite{bey93,sim77}.
In addition, none of the one boson exchange models considered here give
rise to a collective one body effective interaction of the type
used in the earlier analysis \cite{bey89}. So, the origin of the generic
one-body potential, eq.(\ref{UT}), needs clarification.

A generic T-violating P-conserving NN-potential that leads to an
effective one-body potential of the above form has been given by Herzceg
\cite{her66} (isospin dependence suppressed)
\begin{equation}
V_T = \phi_T~ {\bf p}\!\cdot\!{\bf r}~h(r) + h.c.
\label{sigma}
\end{equation}
with an unknown function $h(r)$. In the framework of boson exchanges
this structure is obviously not produced via a simple one boson
exchange. It may, however,  result from two boson exchanges.

Such a potential has been suggested by Huffman\cite{huf70}. A generic
T-violating photon plus meson exchange leads to an effective potential
with definite isospin dependence and calculable $h(r)$, which is
also spin-dependent, if mediated by pion exchange. The resulting
effective one-body potential, defined in eq.(\ref{UT}) has not yet been
calculated.

Another long range force would involve T- and P-odd pion exchange
\cite{sim75} plus some P-odd interaction to compensate for the
P-violation. This would presumably lead to a very tiny effects, since
the P-violating weak coupling is already in the order of $10^{-7}$ and
the P- and T- violating pionic strength through electric dipole moment
measurements is bounded from below through $10^{-10}$ \cite{her88}. The
resulting strength would be of at least $10^{-17}$ or so.

A further mechanism leading to a potential of the type given in
eq.(\ref{sigma}) could be provided through a T-violating P-conserving
interaction of the type given in eqs.(\ref{rho}, \ref{axial}) but
accompanied by a usual strong exchange interaction, not already included
in the wave function. Typically such interactions may occure through
isobar excitations or crossed diagrams, which might then be parametrized
by some effective one boson exchanges similar to the effective
$\sigma$-exchange of strong interactions \cite{bonn}, and lead to a
potential of the type given in eq.(\ref{sigma}). However, all these
rather qualitative statements need a more careful considereration, which
goes beyond the scope of the present paper.

Due to the isospin dependence of $V_T^\rho$ bounds on $\phi_\rho$ may
only be reached in nuclei with protons and neutrons being either only
valence particles or valence holes. Otherwise the T-odd contribution
would be suppressed as is the case for $^{57}$Fe.

Additionally, since none of the potentials, eqs.(\ref{rho}, \ref{axial})
give contributions from closed shells using eq.(\ref{HF}), a better
choice for future $\gamma$ correlation experiments would be such nuclei
with many particles (or holes) outside closed shells, resp. deformed
nuclei.

\end{document}